\documentclass[multphys,vecphys]{styles/svmult}

\usepackage{natbib}

\usepackage{makeidx}         
\usepackage{graphicx}        
\usepackage{multicol}        
\usepackage[bottom]{footmisc}

\makeindex             

\begin{document}

\title*{Cold Gas in Cluster Cores}
\author{Megan Donahue\inst{1}}
\institute{Michigan State University, Physics \& Astronomy Dept., East Lansing, MI US 48824-2320
\texttt{donahue@pa.msu.edu}}
%
%
\maketitle

I review the literature's census of the cold gas in clusters of galaxies. Cold gas here is defined as the gas that
is cooler than X-ray emitting temperatures ($\sim10^7$K) and is not in stars. I present new Spitzer IRAC and MIPS
observations of Abell 2597 that reveal significant amounts of warm dust and star formation at the
level of 5 solar masses per year. 

\section{Introduction: Gas Census in the Core}
\label{sec:Introduction}

Almost immediately following the discovery of clusters as X-ray sources came the 
realization that, as thermal X-ray sources, the gas in the cores of many of these systems 
is radiating energy at a prodigious rate. A simple calculation reveals that the enthalpy ($5nkT/2$ )
content in the gas in such systems would be expended many times during the lifetime 
of the system, unless replenished by source of energy over and above that of gravitational compression. Absent a distributed and significantly large heating source, the gas in this
cooling flow picture should cool, lose pressure support,  and thus 
gently settle deeper into the cluster potential \citep[cf.][]{1994ARA&A..32..277F}.)

The rarity of cool gas and of recently formed stars in the cores of clusters of galaxies
provided the main counter-evidence for the simple cooling flow model in the cores
of clusters throughout the 1990s.  The lack of X-ray coronal lines 
from species such as Fe XVII and O VII 
was the kiss of death for the simple cooling flow model \citep[e.g., ][]{2001A&A...365L.104P}. But that discovery begged 
another question: what supplies the enthalpy radiated by these X-ray sources?
The cold gas in cluster cores provides key clues about the processes 
occurring in the gas there.

Where applicable, I assume that the expansion of the universe can be described with
$H_0=70 h_{70}$ km/s/Mpc.  For ease of terminology, I will call these clusters "cool core"
clusters, although it is possible for the gas density to be high enough for the cluster
to qualify as having a short central cooling time without a temperature gradient. (It is less
awkward than referring to them as "clusters formerly known as cooling flow clusters," in 
any case.)

The core of a cluster of galaxies can usually be defined by the characteristic radius in a beta-model
profile of the form density $\rho(R) \propto (1 + (R/R_c)^2)^{-3\beta/2} $. 
$R_c$ is typically 100-200 $h_{70}^{-1}$ kpc.  Inside this region, the hot gas dominates 
with about $10^{12-13}$~M$_\odot$, while stars provide $\sim10^{12}$~M$_\odot$ of the
mass. The $10^4$K  ionized gas, emitting H$\alpha$, weighs in at about 10$4$  M$_\odot$, and 
as far as we know this gas seems to be always accompanied by collisionally-excited
H$_2$. There can be trace amounts of [OVI]-emitting gas at about $10^6$ K. Some systems may hold significant amounts of cold molecular hydrogen ($10^9- 10^{10}$) M$_\odot$.  
Recent Spitzer observations, reported here and by  \citet[][]{2006ApJ...647..922E} show that these systems also host
star forming regions -- not at the level to explain the old-fashined mass cooling rates, but
enough to produce bright mid-IR Spitzer sources.

I have been invited to discuss the state of knowledge of 
each of these components in the ICM. I note that this area of observational astronomy is
undergoing something of a re-awakening since 2000. The current field is not so much limited by technology,
but by limited application of the technology we have to only a few sources. I am dubious as
to the existence of a single prototypical cool core cluster. I will attempt to show where my conclusions
are based on a detailed observation of a single object, or only on observations of the 
most extreme examples of the class.

\section{Optical Emission Line Nebulae}
\label{sec:optical}

Many of the cool core clusters host prominent optical emission line nebulae with characteristic
radii of about 5-10 kpc. They were first discussed in context with the X-ray clusters by
\citet[][]{1977ApJ...215..723C} and \citet[e.g., ][]{1977MNRAS.180..479F}.  This field has been quite active \citep[][]{1979ApJS...41..147F, 1979ApJ...230..667K, 1981ApJ...250L..59H,
1983ApJ...272...29C, 1985ApJS...59..447H, 1987MNRAS.224...75J, 
 1988ApJ...324L..17R, 1988MNRAS.233..581J,
1989ApJ...338...48H,1992ApJ...397L..31S, 1990ApJ...360L..15V, 1993ApJ...414L..17D,
1995MNRAS.276..947A, 1995MNRAS.274...75C, 
1997ApJ...486..242V, 1999MNRAS.306..857C, 
2006MNRAS.367..433H,  2006MNRAS.371...93W}.
For a review of the early studies (pre-2000), I have chosen to highlight the results in 
 \citet[][]{1989ApJ...338...48H} and \citet[][]{1997ApJ...486..242V} below.
 
The most famous of these nebula, the nebula in NGC1275, \index{NGC1275} spans almost
100 kpc. They exhibit bright, low-ionization optical emission lines such as H$\alpha$,
[OII]3727\AA, [SII]6716/6730\AA, [OI]6300 \AA, and [NII]6548/6584\AA. The total luminosities
are around $10^{41}-10^{43}$ erg s$^{-1}$. They are moderately broadened, at 100-200
km s$^{-1}$. The [SII] lines can be used to derive typical electron densities of 100-300
cm$^{-3}$ \citep[e.g., ][]{1989ApJ...338...48H}, for a variety of sources. 
\citet[][]{1997ApJ...486..242V} studied a single nebula in Abell 2597 \index{A2597} with very
deep optical spectra. With such specta, they could use a combination of the blue and red  
[OII] lines and the [SII] lines (the [OIII] triplet has only provided limits to date) 
\citep[][]{1997ApJ...486..242V}  indicate that
the nebulae are surprisingly hot, $T \sim 10,000 - 12,000$ K. These line ratios indicate
a relatively low photoionization parameter $U\sim 10^{-4}$ \citep[e.g. ][]{1989ApJ...338...48H}, 
where $U$ is the ratio of ionizing photon density to hydrogen particle density.

The line ratio analyses of many of the authors above tend to rule out strong shocks, at least in the
high surface brightness nebulae near the center of the source.  Most of the line
ratios indicate a consistency with an ionization source 
with a black body temperature of  greater than 100,000K, in addition to the expected 
ionization contributions from hot stars.  The lack of He II recombination lines in a deep
spectrum of Abell 2597 \index{A2597} \citep[][]{1997ApJ...486..242V} rules out photons with energies greater than 
54 eV.

The radiative contribution of any AGN to the photoionization and heating  is local at best. Near
an AGN in these sources, one sees enhanced [OIII]/[OII] and prominent peaks in the 
surface brightness of [NII] and [SII]. Work by 
\citet[][]{2005MNRAS.363..216C,2005MNRAS.361...17C}
 has shown that these 
filaments are not strongly turbulent. An interesting result from 
\citet[][]{2006MNRAS.367..433H} regarding
the filaments in NGC1275 \index{NGC1275} that there is an insufficent number of stars to provide the excitation
and heating of the nebular gas. (See Crawford, this proceedings.)

\subsection{2A0335+096, a brief case study of a complicated system}

We have observed the X-ray cluster 2A0335+096. \index{2A0335+096} We show here a new
and deep narrow-band H$\alpha$ image, taken during the early science phase of the SOAR
telescope (a new 4-meter telescope located on Paranal, near Gemini South), together with
re-reduced VLA data, optimizing the high spatial resolution information. We also 
present here a long-slit spectrum that was previously analyzed in only a single spectrum
mode \citep[][]{1993ApJ...414L..17D}. But because the slit was aligned along 
an interesting bar feature in the spectrum, and includes the light from a possibly interacting companion (with its own
emission lines) and extended filament gas, it was worth reviewing the kinematic and 
excitation information (Donahue et al. 2007, submitted to AJ). The radio data show a clear lateral
source, extending perpendicularly to the bar; the H$\alpha$ image, while complex, show
filaments arching around the radio source.

We identifed the two peaks in H$\alpha$ intensity as A and B. These locations are
the approximate peaks in the centers of the main galaxy (A) and the companion galaxy (B), 
separated by about 300 km/s and 4.6 kpc in projection. 
The peak B corresponds to the broadest lines ($\sim 500$ km/s)  
in its region, and also to a region with enhanced [NII]/H$\alpha$ ratios, indicating the presence
of a small AGN in B.
However, the story is different for A. The velocity peak and an abrupt change in the [NII]/H$\alpha$
ratio occurs somewhat south of the surface brightness peak. This location (possibly the
location of the AGN in galaxy A) is also blueshifted compared to A and B.

There is a gradient of gas velocity from A to B, which may be indicative of stripping. 
(The suggestion that it might be tidal is unlikely because of the lack of a similar feature in the 
stars.) From the relative velocities and projected distances, we have estimated that the
companion galaxy/B last interacted with the brightest cluster galaxy (BCG, A) about 60 million
years ago. The radio source timescale is approximately 50 million years. The estimated
star formation rate, from XMM/Optical Monitor UV data and a dust-free H$\alpha$ 
estimate is around 5 solar masses per year. 
We note that the velocity field is not turbulent along the axis of interaction between the
two galaxies. The optical line emission is confined to the region of X-ray emitting gas that
is less than about 2 keV, while the X-ray peak is well-separated from the BCG. 

2A0335+096, \index{2A0335+096} while one of the more complex studied, is probably not all that unusual for
its evidence of a dynamical interaction. What we find interesting here is that radio source
showed evidence for blowing bubbles into the gas in this system and that the timescales
for the interaction and the age of the radio source were quite similar. The complexity
here may be induced by the rearrangement of injection sources while injection was 
occurring.

\subsection{Integral Field Spectroscopy}
2D optical (and IR) spectroscopy is clearly the best way to decompose the  
velocity  fields in these systems. 
\citet[][]{2006MNRAS.371...93W}
have used the Visible Multiobject Spectrograph on the VLT, with 1600 optical fibers to
obtain velocity and line ratio maps for 4 extreme cool-core clusters, Abell 1664, Abell 1835, Abell 2204, 
and Zw 8193. \index{A1664} \index{A1835} \index{A2204} \index{Zw8193} 
The relative velocites of the nebular features in these systems were low,
around 100-300 km/sec. The biggest disturbances were in galaxies with companions.
Remarkably, the [NII]/H$\alpha$ ratio remains relatively constant across the system. It
is possible that the AGN region may not
have been well-sampled by the fibers.

\section{Infrared Emission Line Nebulae}
\label{sec:NIR}

In the early 1990s, when the near-infrared spectroscopic capabilities were becoming
available, \citet[][]{1994iaan.conf..169E}
 began a "cloudy night" project to observe the optical emission
line nebulae of cooling flow clusters. They quickly discovered that these objects were 
bright H$_2$ sources at 2 microns - nearly as luminous as the X-ray emission coming
from the same region. The line ratios of the vibrationally-excited 
 H$_2$ lines indicated excitation temperatures of around 2000K. HST imaging with 
 NICMOS revealed that the morphology of the vibrationally excited H$_2$ emission
 was very similar to that of the optical emission line nebula \citep[][]{2000ApJ...545..670D}.

 Near IR spectroscopy of these objects \citep[][]{
 1997MNRAS.284L...1J, 1998ApJ...494L.155F, 2000MNRAS.318.1232W, 
 2000A&A...354..439K,  2001MNRAS.324..443J, 2002MNRAS.337...49E,
 2002MNRAS.337...63W}
  showed that the spectrum was consistent with heating by a hot $T>50,000$K 
 stellar continuum, electron densities of over $10^5$ cm$^{-3}$, and line ratios inconsistent
 with that of shocks. Therefore this gas is over-pressured by a factor of 
 about 100-1000 compared to the pressures derived for the optical and X-ray gas. 
 
\citet[][]{2005MNRAS.360..748J}
 detect near IR line emission out to 20 kpc from the centers
 of these systems, gas dynamics that are highly coupled with that of the optical emission-line
 gas, and also find evidence for an ionizing source consistent with a black body temperature
 of over 100,000 K. 
 
 I find it extremely interesting that the "Mystery Ionization Sources'' for both the optical
 filaments and the near infrared-emitting molecular gas have the similar property that
 both require EUV radiation, yet these photons must be less energetic than that required
 to produce He II recombination emission \citep[][]{1997ApJ...486..242V}. More He II
 searches are needed to see if that deficit in the optical spectrum is common. 
 
 \section{Cold Molecular Hydrogen and CO}
\label{sec:cold}

The decade of the 90s experienced many searches for the cold molecular gas associated
with the putative cooling flow
\citep[][]{ 1990ApJ...355..401G, 1994A&A...281..673M, 1994ApJ...422..467O, 
1994A&A...283..407B, 1995AJ....109...26O, 2000PASJ...52..743F}. However, it wasn't until the
technology had progressed such that significant detections and maps were produced
\citep[][]{2001MNRAS.328..762E, 
2003ApJ...594L..13E,
2003A&A...412..657S,
2004A&A...415L...1S,
2006A&A...454..437S}
 These
detections were achieved with the JCMT and the IRAM. They saw between 300 million and
40 billion solar masses of cold ($T \sim 30$K) molecular hydrogen (inferred from the presence of CO) 
in the $>10^{42}$ erg/s H$\alpha$-luminous systems that they
targeted. Their successful detections were
most likely in systems with the brightest H$\alpha$ and 2-micron H$_2$ lines.
 
  \section{Hot Gas - [OVI] Line Emission}
\label{sec:hot}

The existence of the [OVI] emission line doublet at 1032/1035 \AA\AA indicates the presence
of million K gas. Non-primordial gas at this temperature cools extremely rapidly.  It was found most commonly
in our own Galaxy, first by the Copernicus satellite and more recently by FUSE. 
FUSE made long, night-time spectral
observations of the central galaxies of several clusters. 
\citet[][]{2001ApJ...560..187O} reported a 
detection in the cluster Abell 2597, but not Abell 1795. \index{A2597} \index{A1795} Later,
\citet[][]{2004A&A...421..503L}
reported only upper limits in Abell 2029 and Abell 3112, \index{A2029} \index{A3112} while most recently, 
\citet[][]{2006ApJ...642..746B} reported detections in the re-analyzed Abell 1795  \index{A1795} data and
in NGC1275, \index{NGC1275} but not in the group AWM7. \index{AWM7}

So far, the detections are only in systems with known optical line emission. None of the reported 
detections are extremely strong. The  \citet[][]{2001ApJ...560..187O} 
Abell 2597 \index{A2597} detection is of only one member of the 
doublet (the fainter one is obscured), while the Abell 1795 \index{A1795} re-analysis by
 shows \citet[][]{2006ApJ...642..746B}
excess in two locations, at the quoted $3-\sigma$ level, but cut up by H$_2$ and FeII 
Galactic absorption, and in the presence of significant continuum. It is a very difficult experiment on a
small telescope, and appropriate caution should be applied. (The author can say there
may be lessons learned from the Einsteint Observatory FPCS spectrum of NGC1275, \index{NGC1275} 
\citep[][]{1988cfcg.work...63C}, a
spectrum obtained with a 1 cm$^{2}$ effective collecting area, reporting putative Fe XVII 
lines in the presence of X-ray continuum light.)
That said, these limits and detections of [OVI] are consistent with a rather large cooling rate in the 
central region of the clusters, of about 30 solar masses per year in Abell 1795. \index{A1795} Such a 
rate, if turned into stars would easily be visible to Spitzer observations.

\section{Dust and PAHs: Spitzer's Infrared Vision}
\label{sec:dust}

While BCGs are not exactly know for their dust emission, even single-orbit HST images show
that these systems have dust lane (e.g. A2052, A2597, PKS0745-191, 2A0335+096,
A4059), \index{A2052} \index{A2597} \index{PKS0745-19} \index{2A0335+096}
as noted by \citet[][]{
2004ApJ...606..185C, 2000ApJ...545..670D, 1996ApJ...466L...9M}, this presentation. 
Older groundbased extinction maps of  NGC4696 \index{NGC4969}  \citep[][]{1989ApJ...345..153S}
showed evidence for dust in that system that followed the emission-line gas. However,
this detailed correspondence between dust and emission-line does not usually exist 
(e.g., Donahue et al. 2000).

Extinction towards the optical emission line filaments have been estimated from the
decrement in a Balmer line series to be $A_V  \sim 1$ 
\citep[][]{1997ApJ...486..242V}.
IUE-based Lyman-$\alpha$ to ground-based H$\alpha$ ratios 
\citep[][]{1992ApJ...391..608H} suggest
$E(B-V) \sim 0.2$. 
\begin{figure}
\centering
\includegraphics[height=4cm]{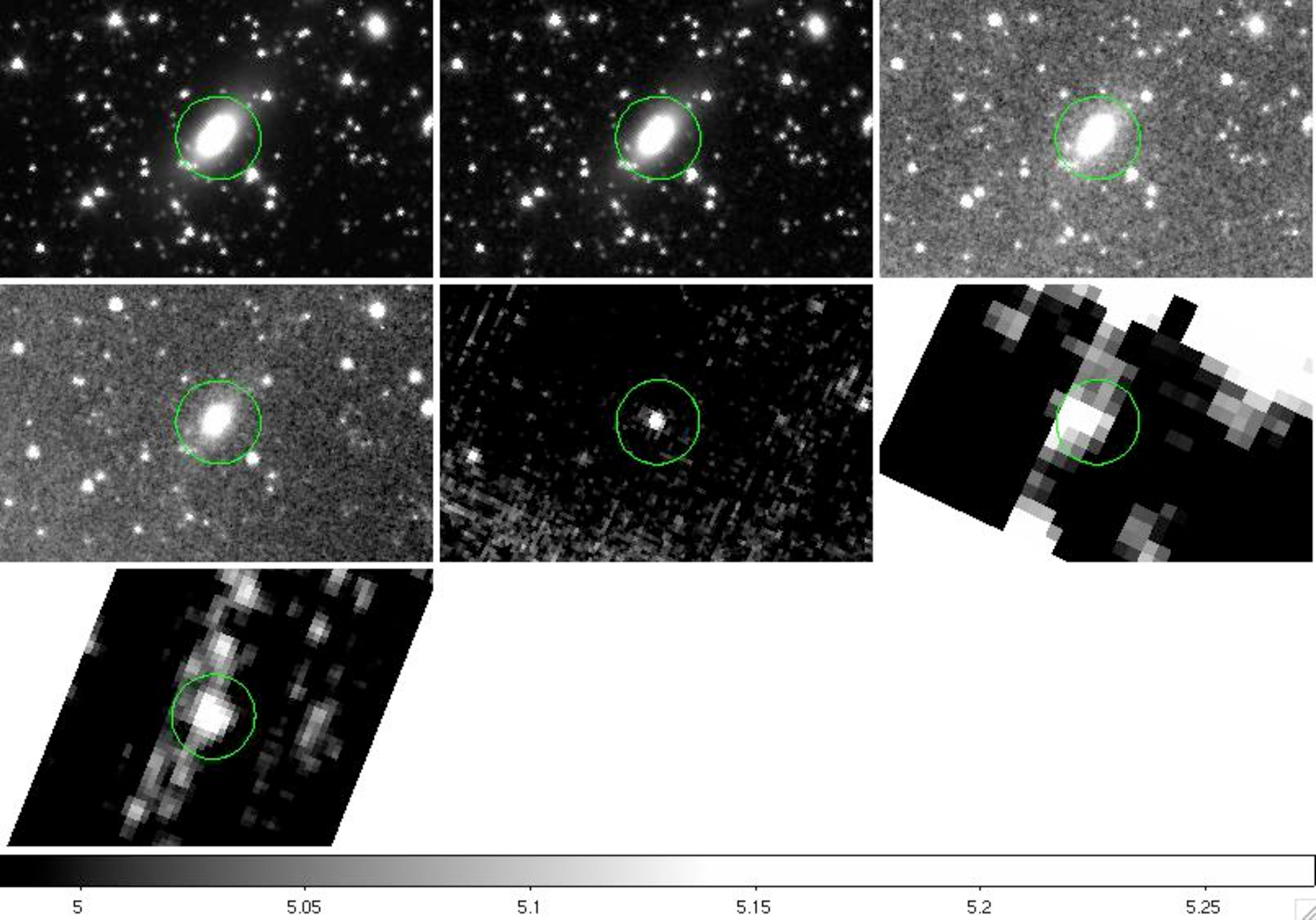}
%
%
\caption{Images of the BCG in A2597 \index{A2597} by Spitzer. The numbers in the corners of the
images corresponds to the central wavelength of each bandpass, in microns. }
\label{fig:A2597pix}       
\end{figure}

\begin{figure}
\centering
\includegraphics[height=4cm]{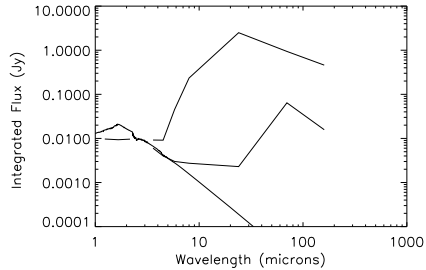}
%
%
\caption{Spitzer SED for the BCG in Abell 2597. \index{A2597} 
The upper curve is the full flux in the aperture; the middle curve is
the background-corrected flux; the lower curve is the expectation based on stellar spectra
alone (a RJ tail). The 2MASS photometry points are from the extended source catalog. (Donahue 
\& Sparks 2007, in prep.)}
\label{fig:A2597SED}       
\end{figure}

\citet[][]{1993ApJ...414L..17D}
showed that the emission-line gas {\em itself} is dusty, based on the
intensity of calcium emission lines. The lack of calcium lines from the ionized gas 
shows that calcium has been strongly depleted into grains.

Dust emission was difficult to find, pre-Spitzer. IRAS had poor spatial resolution and low sensitivity, which
resulted in only upper limits for dust in clusters. ISO had a scan mode, but here too the low
spatial resolution and the low sensitivity limits the constraints. 
\citet[][]{2002A&A...383..367S}  found excesses
and deficits of emission at 120 and 180 microns towards 6 nearby clusters. One promising
lead may
be the development of SCUBA, 
\citet[][]{1999MNRAS.306..599E}
 and, at $14"$ resolution,  
\citet[][]{2002MNRAS.330...92C}, report 2 detections out of 7 clusters observed at 850 microns. NGC1275
\index{NGC1275}
has been detected by SCUBA
\citep[][]{2001MNRAS.328..359I}. They report about 60 
million solar masses of extended, 20K, dust in NGC1275.

\citet[][]{2006ApJ...647..922E}
report Spitzer mid-IR (70 micron) detections of 3 clusters, Abell 2390, \index{A2390}
Zw 3146, \index{Zw3146} and Abell 1835, \index{A1835} in the course of their search for high-redshift lensed galaxies
behind moderate redshift clusters of galaxies. They picked the 3 BCGs in their sample with
the brightest MIPS 24-micron detections to follow up with longer wavelength 
observations.

We present here new Spitzer data for Abell 2597 
\index{A2597} (Donahue \& Sparks 2007, in prep). The BCG in Abell 2597 is well-detected
in both IRAC (3.6, 4.5, 5.8, and 8 microns) and MIPS (24, 70, and 160 micron) observations.
The total luminosity in the far infrared is approximately $10^{44}$ erg s$^{-1}$. The broad-band
spectral energy distribution shows an excess at 70 and 160 microns that well-exceeds the
expected contribution from the Rayleigh-Jeans tail of the stellar contribution. 
(Figures \ref{fig:A2597pix}  and \ref{fig:A2597SED}.) Furthermore, the excesses at 5.8 and 8
microns indicate the presence of strong PAH emission, which should be confirmed by IRS
spectroscopy once the extended source analysis is understood.

These observations consisted of 270 minutes in the IRAC band and only 36 minutes
in the MIPS. The far infrared luminosity corresponds to about 4 solar masses per year 
\citep[][]{1998ApJ...498..541K}, which
is inconsistent with the [OVI] cooling luminosity of $20\pm5$ M$_\odot$ yr$^{-1}$ 
\citep[][]{2001ApJ...560..187O}. The inferred star 
formation rate is consistent with the UV star formation rate inferred from HST/STIS 
UV observations by O'Dea et al. (2004) and with the H$\alpha$ star
formation rate within about $r=2"$ 
\citep[][]{2000ApJ...545..670D} , 
 about 2 solar masses per  year based on the relations described in 
\citet[][]{1998ApJ...498..541K}.

\section{Conclusions}
\label{sec:Conclusions}

We have reported a major Spitzer result for BCGs in cool core clusters of galaxies: 
a 70-160 micron excess well above the Rayleigh-Jeans tail of a stellar contribution. The
luminosity and the spectrum are typical of a Luminous InfraRed Galaxy (LIRG) or
a starburst. These spectra are consistent with UV and optical indicators of
star formation in Abell 2597. 

The various cold constituents of the clusters of galaxies reviewed here 
suggest a common excitation mechanism
for all phases, at least for the brightest parts. The evidence for this includes: common
hosts, similar photoionization and heating constraints, a relative lack of line ratio
variation across the nebulae. However, these issues
are far from settled, and may provide vital clues as to the mechanisms for cooling gas,
feedback, and star formation in the local universe, in the largest galaxies known.

The questions remaining include: Are galaxy-galaxy interactions required to stimulate
central optical / infrared nebulae? Are optical filaments trails of galaxies punching through
a molecular hydrogen reservoir as suggested by Wilman et al. (2006), or are they
lofted by buoyant radio plasma as suggested by Crawford and Fabian? What is 
the relationship between nebulae and star formation? Are the filaments farther from
the stars excited by different mechanism from the filaments close in? What is
this mysterious FUV (but not too FUV) ionization source for the optical and infrared
filaments?

What next? I would like to see more Spitzer MIPS and MIPS-SED observations done
before the cryocoolant is exhausted, at the end of April 2009 and the end of a short 
Cycle 5. These spectra and images are needed to get crtiical dust temperatures (see
O'Dea cycle 3 program, with many co-Is in the community.) It would be very good to 
trace the velocity fields of the ICM using IFU and efficient (mapping mode) long-slit
observations. Extremely deep optical spectra can provide emission-line diagnosticis
to get nearly model-independent measures of T, Z, n, U (e.g. Voit \& Donahue 1997).
Deep X-ray observations should eventually reveal the faint coronal lines that ought
to be present if some cooling gas is fueling the observed star formation. Finally, probing
the relationship between the line-emitting gas, the dust emission, and the low-entropy
X-ray emitting gas will require multi-wavelength imaging at somewhat similar
spatial resolutions in order to test whether the morphologies of the emission regions
really are similar.

%
%
%

%
%
 \bibliographystyle{apj}
 \bibliography{donahue_reference}
%


\printindex
\end{document}